\documentclass[a4paper,11pt]{article}
\usepackage{jinstpub}
\usepackage{microtype}
\usepackage{booktabs}
\usepackage{subcaption}
\usepackage{float}
\usepackage{arydshln}

\title{Dependence of polytetrafluoroethylene reflectance on thickness at visible and ultraviolet wavelengths in air}

\collaboration{The NEXT Collaboration}
\author[11,a]{S.~Ghosh,\note[a]{Corresponding author.}}
\author[11,a]{J.~Haefner}
\author[11,b]{J.~Mart\'in-Albo,\note[b]{Now at the Instituto de F\'isica Corpuscular (IFIC), Valencia, Spain.}}
\author[11]{R.~Guenette,}
\author[11]{X.~Li,}
\author[11,c]{A.A.~Loya Villalpando,\note[c]{Now at the California Institute of Technology, Pasadena, CA, United States.}}
\author[11]{C.~Burch,}
\author[2]{C.~Adams,}
\author[22]{V.~\'Alvarez,}
\author[6]{L.~Arazi,}
\author[20]{I.J.~Arnquist,}
\author[4]{C.D.R~Azevedo,}
\author[2]{K.~Bailey,}
\author[22]{F.~Ballester,}
\author[16]{J.M.~Benlloch-Rodr\'{i}guez,}
\author[14]{F.I.G.M.~Borges,}
\author[3]{N.~Byrnes,}
\author[19]{S.~C\'arcel,}
\author[19]{J.V.~Carri\'on,}
\author[23]{S.~Cebri\'an,}
\author[20]{E.~Church,}
\author[14]{C.A.N.~Conde,}
\author[11]{T.~Contreras,}
\author[21]{G.~D\'iaz,}
\author[19]{J.~D\'iaz,}
\author[5]{M.~Diesburg,}
\author[14]{J.~Escada,}
\author[22]{R.~Esteve,}
\author[6,7,19]{R.~Felkai,}
\author[13]{A.F.M.~Fernandes,}
\author[13]{L.M.P.~Fernandes,}
\author[16,9]{P.~Ferrario,}
\author[4]{A.L.~Ferreira,}
\author[13]{E.D.C.~Freitas,}
\author[8]{A.~Goldschmidt,}
\author[16,9,d]{J.J.~G\'omez-Cadenas,\note[d]{NEXT Collaboration's co-spokesperson.}}
\author[21]{D.~Gonz\'alez-D\'iaz,}
\author[10]{R.M.~Guti\'errez,}
\author[2]{K.~Hafidi,}
\author[1]{J.~Hauptman,}
\author[13]{C.A.O.~Henriques,}
\author[21]{J.A.~Hernando~Morata,}
\author[16]{P.~Herrero,}
\author[22]{V.~Herrero,}
\author[6,7]{Y.~Ifergan,}
\author[3]{B.J.P.~Jones,}
\author[21]{M.~Kekic,}
\author[18]{L.~Labarga,}
\author[3]{A.~Laing,}
\author[5]{P.~Lebrun,}
\author[19]{N.~L\'opez-March,}
\author[10]{M.~Losada,}
\author[13]{R.D.P.~Mano,}
\author[16]{A.~Mart\'inez,}
\author[19]{M.~Mart\'inez-Vara,}
\author[19,21,e]{G.~Mart\'inez-Lema,\note[e]{Now at the Weizmann Institute of Science, Rehovot, Israel.}}
\author[3]{A.D.~McDonald,}
\author[16,9]{F.~Monrabal,}
\author[13]{C.M.B.~Monteiro,}
\author[22]{F.J.~Mora,}
\author[19]{J.~Mu\~noz Vidal,}
\author[19]{P.~Novella,}
\author[3,c]{D.R.~Nygren,}
\author[21,19]{B.~Palmeiro,}
\author[5]{A.~Para,}
\author[12]{J.~P\'erez,}
\author[19]{M.~Querol,}
\author[6]{A.~Redwine,}
\author[21]{J.~Renner,}
\author[2]{J.~Repond,}
\author[2]{S.~Riordan,}
\author[17]{L.~Ripoll,}
\author[10]{Y.~Rodr\'iguez Garc\'ia,}
\author[22]{J.~Rodr\'iguez,}
\author[3]{L.~Rogers,}
\author[16,12]{B.~Romeo,}
\author[19]{C.~Romo-Luque,}
\author[14]{F.P.~Santos,}
\author[13]{J.M.F. dos~Santos,}
\author[6]{A.~Sim\'on,}
\author[19]{M.~Sorel,}
\author[15]{T.~Stiegler,}
\author[22]{J.F.~Toledo,}
\author[16]{J.~Torrent,}
\author[19]{A.~Us\'on,}
\author[4]{J.F.C.A.~Veloso,}
\author[15]{R.~Webb,}
\author[6,f]{R.~Weiss-Babai,\note[f]{On leave from Soreq Nuclear Research Center, Yavneh, Israel.}}
\author[15,g]{J.T.~White,\note[g]{Deceased.}}
\author[3]{K.~Woodruff,}
\author[19]{N.~Yahlali}
\affiliation[1]{
Department of Physics and Astronomy, Iowa State University, 12 Physics Hall, Ames, IA 50011-3160, USA}
\affiliation[2]{
Argonne National Laboratory, Argonne, IL 60439, USA}
\affiliation[3]{
Department of Physics, University of Texas at Arlington, Arlington, TX 76019, USA}
\affiliation[4]{
Institute of Nanostructures, Nanomodelling and Nanofabrication (i3N), Universidade de Aveiro, Campus de Santiago, Aveiro, 3810-193, Portugal}
\affiliation[5]{
Fermi National Accelerator Laboratory, Batavia, IL 60510, USA}
\affiliation[6]{
Nuclear Engineering Unit, Faculty of Engineering Sciences, Ben-Gurion University of the Negev, P.O.B. 653, Beer-Sheva, 8410501, Israel}
\affiliation[7]{
Nuclear Research Center Negev, Beer-Sheva, 84190, Israel}
\affiliation[8]{
Lawrence Berkeley National Laboratory (LBNL), 1 Cyclotron Road, Berkeley, CA 94720, USA}
\affiliation[9]{
Ikerbasque, Basque Foundation for Science, Bilbao, E-48013, Spain}
\affiliation[10]{
Centro de Investigaci\'on en Ciencias B\'asicas y Aplicadas, Universidad Antonio Nari\~no, Sede Circunvalar, Carretera 3 Este No.\ 47 A-15, Bogot\'a, Colombia}
\affiliation[11]{
Department of Physics, Harvard University, Cambridge, MA 02138, USA}
\affiliation[12]{
Laboratorio Subterr\'aneo de Canfranc, Paseo de los Ayerbe s/n, Canfranc Estaci\'on, E-22880, Spain}
\affiliation[13]{
LIBPhys, Physics Department, University of Coimbra, Rua Larga, Coimbra, 3004-516, Portugal}
\affiliation[14]{
LIP, Department of Physics, University of Coimbra, Coimbra, 3004-516, Portugal}
\affiliation[15]{
Department of Physics and Astronomy, Texas A\&M University, College Station, TX 77843-4242, USA}
\affiliation[16]{
Donostia International Physics Center (DIPC), Paseo Manuel Lardizabal, 4, Donostia-San Sebastian, E-20018, Spain}
\affiliation[17]{
Escola Polit\`ecnica Superior, Universitat de Girona, Av.~Montilivi, s/n, Girona, E-17071, Spain}
\affiliation[18]{
Departamento de F\'isica Te\'orica, Universidad Aut\'onoma de Madrid, Campus de Cantoblanco, Madrid, E-28049, Spain}
\affiliation[19]{
Instituto de F\'isica Corpuscular (IFIC), CSIC \& Universitat de Val\`encia, Calle Catedr\'atico Jos\'e Beltr\'an, 2, Paterna, E-46980, Spain}
\affiliation[20]{
Pacific Northwest National Laboratory (PNNL), Richland, WA 99352, USA}
\affiliation[21]{
Instituto Gallego de F\'isica de Altas Energ\'ias, Univ.\ de Santiago de Compostela, Campus sur, R\'ua Xos\'e Mar\'ia Su\'arez N\'u\~nez, s/n, Santiago de Compostela, E-15782, Spain}
\affiliation[22]{
Instituto de Instrumentaci\'on para Imagen Molecular (I3M), Centro Mixto CSIC - Universitat Polit\`ecnica de Val\`encia, Camino de Vera s/n, Valencia, E-46022, Spain}
\affiliation[23]{
Centro de Astropart\'iculas y F\'isica de Altas Energ\'ias (CAPA), Universidad de Zaragoza, Calle Pedro Cerbuna, 12, Zaragoza, E-50009, Spain}
\emailAdd{soumya\_ghosh@g.harvard.edu}
\emailAdd{jhaefner@g.harvard.edu}

\abstract{Polytetrafluoroethylene (PTFE) is an excellent diffuse reflector widely used in light collection systems for particle physics experiments. However, the reflectance of PTFE is a function of its thickness. In this work, we investigate this dependence in air for light of wavelengths 260~nm and 450~nm using two complementary methods. We find that PTFE reflectance for thicknesses from 5~mm to 10~mm ranges from 92.5\% to 94.5\% at 450~nm, and from 90.0\% to 92.0\% at 260~nm. We also see that the reflectance of PTFE of a given thickness can vary by as much as 2.7\% within the same piece of material. Finally, we show that placing a specular reflector behind the PTFE can recover the loss of reflectance in the visible without introducing a specular component in the reflectance.}

\keywords{Time projection chambers}

\arxivnumber{1234.56789} 

\begin{document}
\maketitle
\flushbottom

\section{Introduction} \label{sec:Introduction}
Polytetrafluoroethylene (PTFE), often referred to by the brand name \textsc{Teflon}, is a commonly used plastic in particle physics experiments (see, for example, Ref.~\cite{Alvarez:2012sma, Auger:2012gs, Akerib:2015cja}) due to its excellent diffuse reflectance and affordability. The NEXT experiment \cite{Martin-Albo:2015rhw}, designed to search for neutrinoless double beta decay, uses PTFE as a diffuse reflector around the drift volume of the high pressure gaseous xenon time projection chamber (TPC) to improve light collection. The measurements described below will inform the design decisions for the upcoming phases of NEXT, including NEXT-100 \cite{Alvarez:2012sma, Martin-Albo:2015rhw} and a future tonne-scale experiment \cite{Adams:2020cye}.

While PTFE is excellent for light collection, its radiopurity can be a concern for low-background experiments \cite{Novella:2019cne}, especially at large scale where a significant quantity of material is needed. Moreover, PTFE can absorb gaseous xenon \cite{Rogers:2018lle}, leading to a loss of active detector volume. Minimizing the amount of PTFE in such experiments is therefore an attractive solution, provided that the light collection remains high.

Noble gases scintillate primarily in the vacuum ultraviolet (VUV) range of the electromagnetic spectrum, and some investigations of PTFE reflectance in this region have been done in the past. Reflectance of PTFE at 175~nm at a single 5~mm thickness was studied thoroughly in Ref.~\cite{Silva:2009ip}, and reflectance in liquid xenon for thicknesses from 1~mm to 9.5~mm was investigated in Ref.~\cite{Haefner:2016ncn}. Further, reflectance as a function of angle was studied in Ref.~\cite{Kravitz:2020}. However, as many experiments (including NEXT) use tetraphenyl butadiene (TPB) wavelength shifter to shift the 175~nm scintillation light of xenon to the visible ($\sim420$~nm), studies of reflectance as a function of PTFE thickness in the visible are also of high interest. While it is generally accepted that PTFE has around 99\% reflectance \cite{Silva:2009ip} in the visible, information on its variation as a function of thickness is also relevant.

In this work, the reflectance variation of PTFE with thickness is investigated with two different methods. The first method uses an integrating sphere with a \textsc{Cary 7000 Universal Measurement Spectrophotometer}. The integrating sphere is an Agilent Diffuse Reflectance Accessory\footnote{https://www.agilent.com/cs/library/flyers/public/5991-1717EN\_PromoFlyer\_UV\_DRA.pdf}. The second method involves enclosed boxes made of PTFE of different thicknesses where the reflectance is studied by measuring the signal as a function of distance from the end of the box. The measurements presented here are made for 260~nm and 450~nm.

\section{Measurements} \label{sec:Measurements}
In order to investigate the reflectance of PTFE, several $12~\mathrm{in}\times24~\mathrm{in}$ sheets of different thicknesses were purchased. All sheets for the main reflectance measurements were sourced from \textsc{ePlastics}\footnote{www.eplastics.com}. The thicknesses used are 4.76~mm, 6.53~mm, 7.94~mm, and 9.53~mm. For brevity, we refer to these, respectively, as 5~mm, 6~mm, 8~mm, and 10~mm.

The pieces cut from the PTFE sheets were all treated and cleaned in the same way. One side of each piece was sanded by hand with 300, then 1000, and finally 2000 grit sandpaper to make the surface uniform and matte. Each piece was then cleaned in an ultrasonic bath in a solution of deionized (DI) water and Alconox for 15 minutes, finally followed with an ultrasonic bath in pure DI water for 15 minutes. The pieces were air-dried overnight before measurement. All cleaning was completed in a single day, each step immediately following the last. Pieces were generally cleaned and measured within a few days of being cut, although sometimes remeasured several days or weeks later. Pieces were  stored in a cabinet wrapped in tin foil. This is comparable to cleaning methods used in particle physics experiments (cf.~\cite{Haefner:2016ncn}). Other surface treatments for radon removal will be explored in future works.

\subsection{Method 1: Spectrophotometer} \label{sec:method1}
\begin{figure}[tb]
\centering
\includegraphics[width=0.49\textwidth]{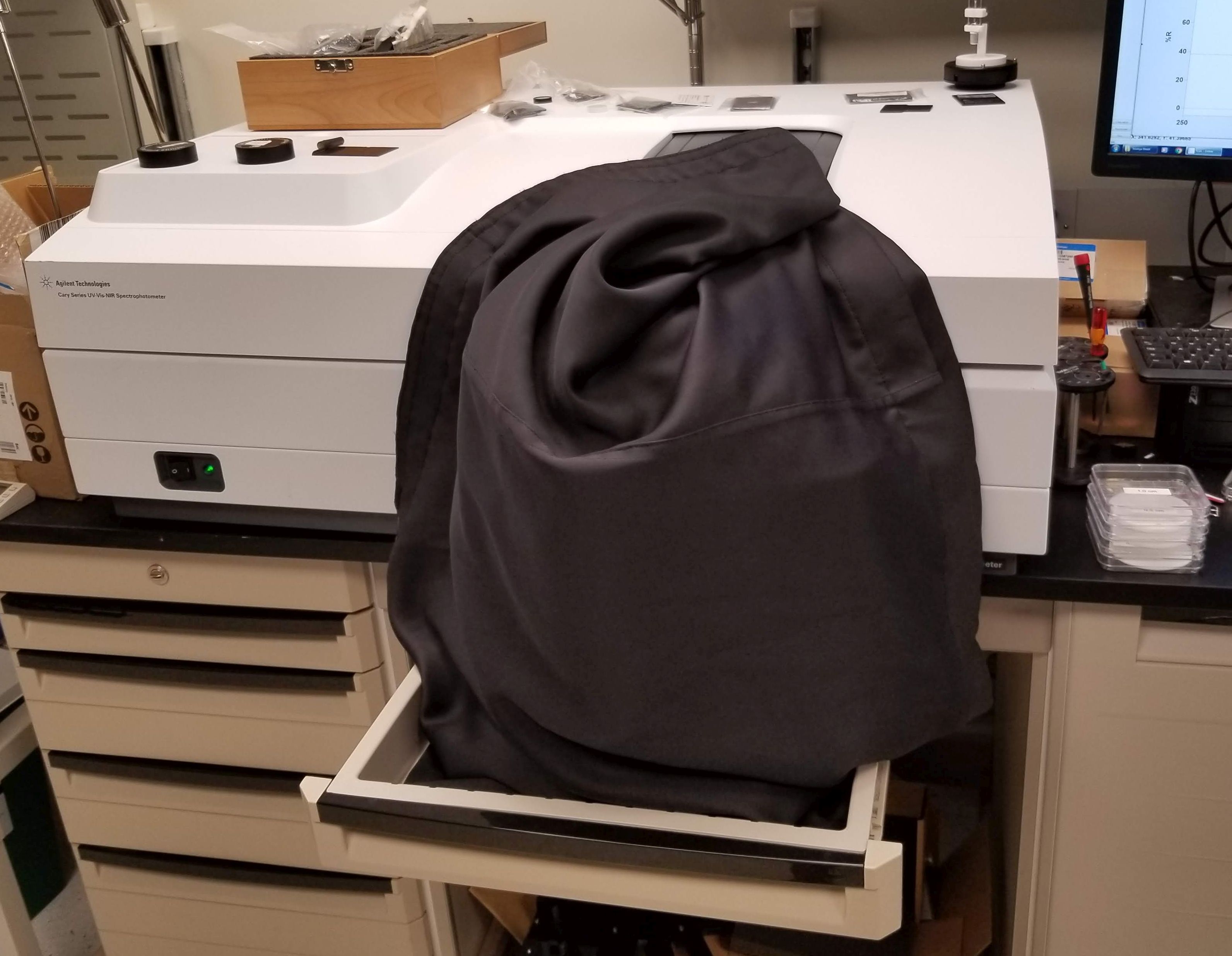}
\includegraphics[width=.49\textwidth]{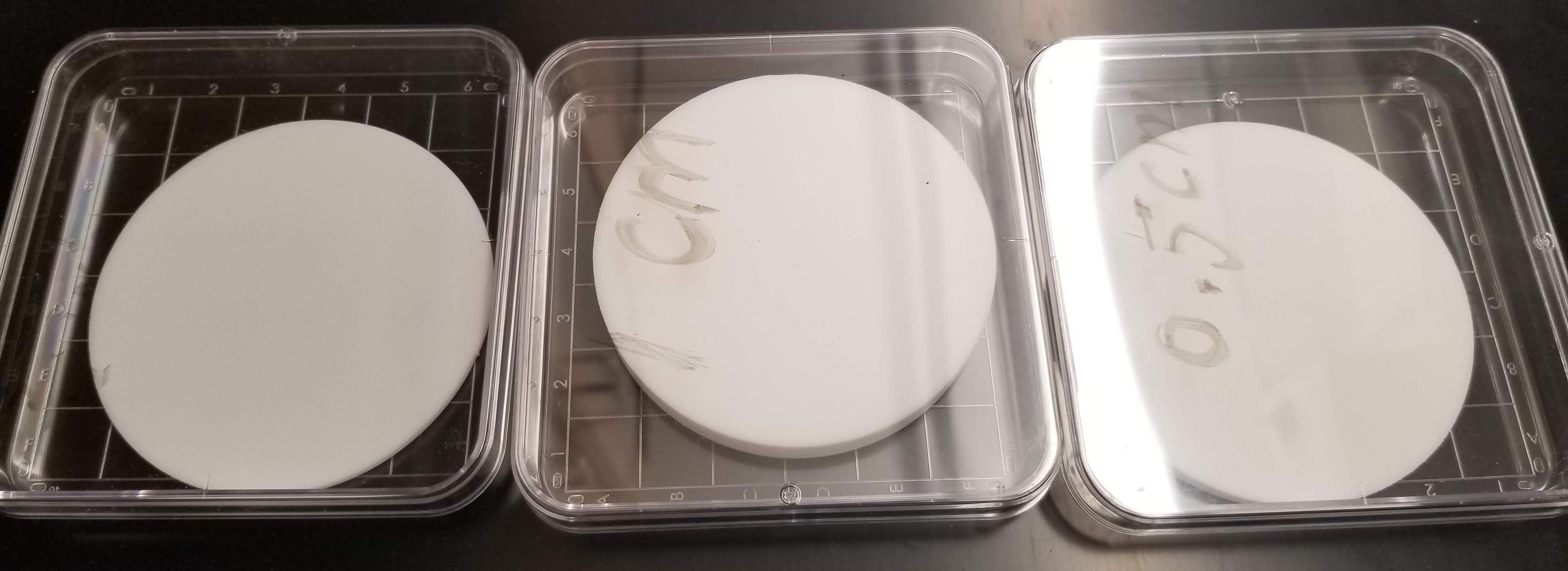}
\caption{Left: The \textsc{Cary 7000 spectrophotometer} of the Harvard Center for Nanoscale Systems (CNS) during a measurement. The compartment containing the integrating sphere is covered with a blackout shroud to prevent light leaks. The PTFE disks used to assess systematic variation can be seen in the transparent sample boxes on the right. Right: A closer look at the PTFE disks.}
\label{fig:photometer_with_shroud}
\end{figure}

The first method for determining reflectance makes use of a spectrophotometer (SPM), the \textsc{Agilent Cary 7000 Universal Measurement Spectrophotometer} (Figure~\ref{fig:photometer_with_shroud}), owned and maintained by the Harvard Center for Nanoscale Systems (CNS). The SPM is coupled to a reflecting sphere and allows measurement of the reflectance of a given sample \textit{relative} to some reference, as a function of wavelength.

\begin{figure}[!htb]
\centering
\includegraphics[width=0.7\textwidth]{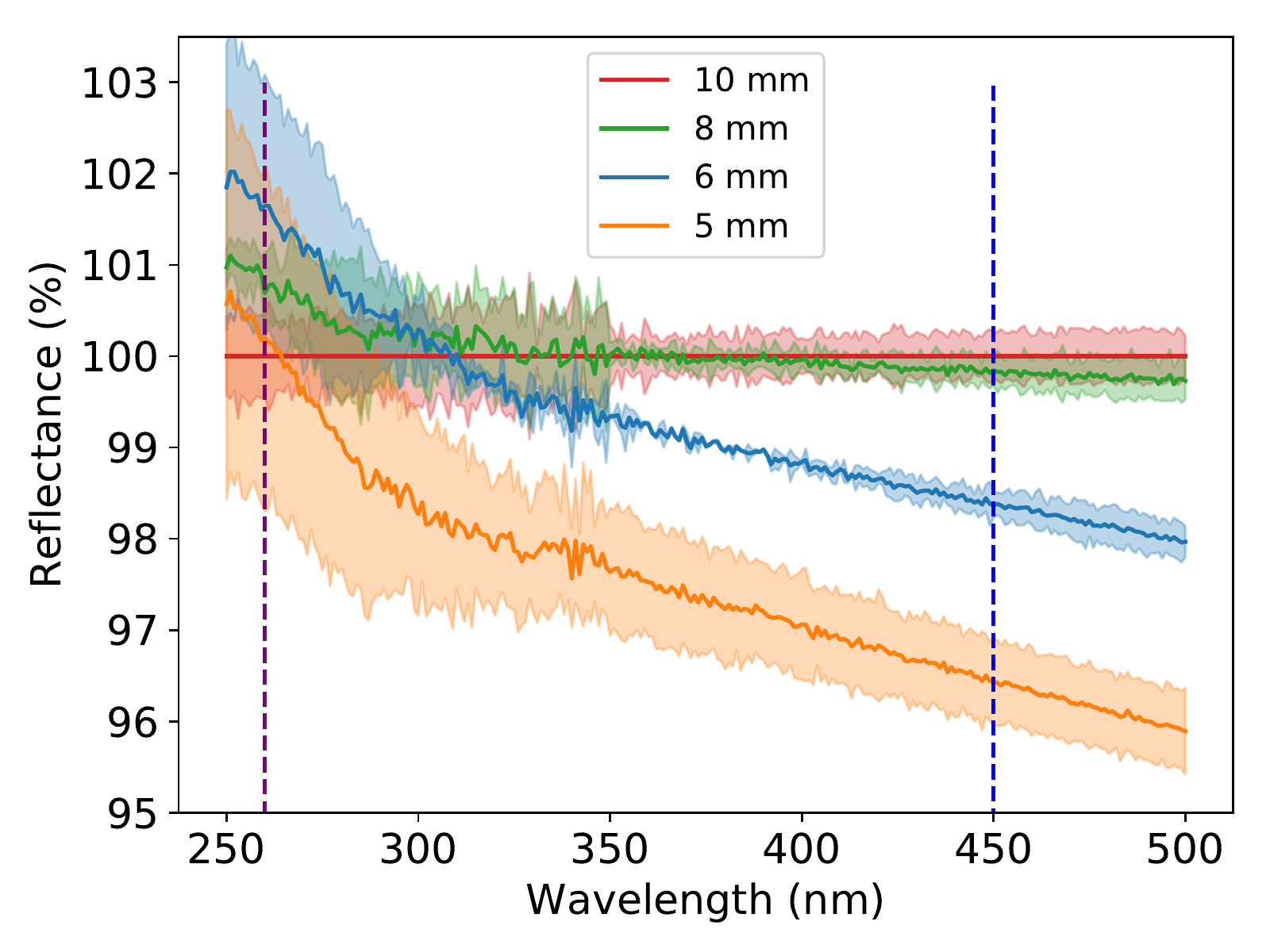}
\caption{Example of reflectance measurements from the spectrophotometer for different PTFE thicknesses relative to the 10-mm sample. The purple and blue dashed lines represent the values of the LED used in method 2 (260~nm and 450~nm, respectively). Determination of values and uncertainties is described in Section~\ref{sec:spm_results_section}. The differences only rise to the few percent level at most.}
\label{fig:spm}
\end{figure}

\subsubsection{Spectrophotometer setup}
The primary measurement in the spectrophotometer comes from averaging the reflectance of two pieces cut from two different locations on the same PTFE sheet. These are in fact the same pieces that are used in the measurements described in Section~\ref{sec:ptfe_boxes}, which are carried out on boxes built out of PTFE. As such the pieces are referred to as the "box back" and the "box side," as they are the respective sides and backs of boxes of a given thickness. Our error bars consisted of a component estimating the variance measurement to measurement of the same piece, and variation between different pieces cut from the same sheet, added in quadrature. The variation between different pieces came from the variance of the box back versus box side measurements. To assess variations between repeated measurements, 3-inch diameter disks (visible on the right side of Fig.~\ref{fig:photometer_with_shroud}) were cut from the PTFE sheets of different thicknesses and measured three times with the SPM. A difference of up to 2.7\% (relative to 1 cm reference) is observed when studying the reflectance of different pieces from the same sheet.

The SPM measurements of the reflectance, relative to the reference sample of 10~mm, for the different thicknesses are shown in Figure~\ref{fig:spm}. The error bars are calculated as described previously.

\subsubsection{Results from the spectrophotometer}
\label{sec:spm_results_section}
A summary of the different reflectances measured with the SPM is presented in Table~\ref{tab:spm_main_result}. The results are shown for 260 and 450~nm in order to directly compare with the second method (see Section~\ref{sec:comparison}), where LEDs of these wavelengths are used.

The data points are the average of single measurements of box back and side panel pieces. The uncertainties come from the standard deviation of triplicate measurements of PTFE disks of the corresponding thicknesses, added in quadrature to the standard deviation between the box side and back measurements. The variation between subsequent disk measurements is generally around 0.1\%; the difference between box back and side is by far the dominant contribution. Please note that different pieces are used to obtain the mean of the data points versus the error bars. The disks provide our estimate of the variation between subsequent measurements of the same sample, but they are \textit{not} the pieces whose actual reflectances are reported, only their standard deviation. The reason for not averaging in the means of the disk measurements is that degradation moved them far from the box reflectances, see Section~\ref{sec:degrade}.

At 260~nm, the reflectance of all thicknesses measured is the same within errors, indicating that the UV reflectance of PTFE is not very dependent on the thickness of the material. At 450~nm, between 5 and 10~mm we measure a difference of 3.6 $\pm$ 0.6\% in reflectance.

\begin{table}[!htb]
\centering
\caption{Relative reflectance (with respect to the reference sample of 10~mm) of PTFE box pieces for various thicknesses at 260~nm and 450~nm measured using the spectrophotometer. Compare to Figure~\ref{fig:spm}. See Section~\ref{sec:spm_results_section} for specifics of value and error bar determination.} \label{tab:spm_main_result}
\begin{tabular}{ccc} 
\toprule
Thickness & 260 nm & 450 nm \\ 
\midrule
10 mm & 100 $\pm$ 0.4\% & 100 $\pm$ 0.3\% \\ 
8 mm & 100.7 $\pm$ 0.5\% & 99.8 $\pm$ 0.2\% \\ 
6 mm & 101.6 $\pm$ 1.5\% & 98.4 $\pm$ 0.2\% \\
5 mm & 100.2 $\pm$ 1.9\% & 96.4 $\pm$ 0.5\% \\
\bottomrule
\end{tabular}
\end{table}

\subsection{Method 2: PTFE boxes}
\label{sec:ptfe_boxes}
The two significant disadvantages of the SPM method described in Section~\ref{sec:method1} are that the measurements are relative and that the small area of the piece that is measured (approximately 0.2~cm$^2$) makes the measurements highly sensitive to potential reflectance variation across the piece or the sheet from which they were cut. The alternative method described in this section, referred to as the \emph{box method}, uses a PTFE cuboid of a total surface area of 749~cm$^2$. This method provides an absolute reflectance and averages potential reflectance variations within the PTFE samples, as it covers a much larger surface area.

\subsubsection{Box method setup}
\begin{figure}[tb]
\centering
\includegraphics[width=0.7\textwidth]{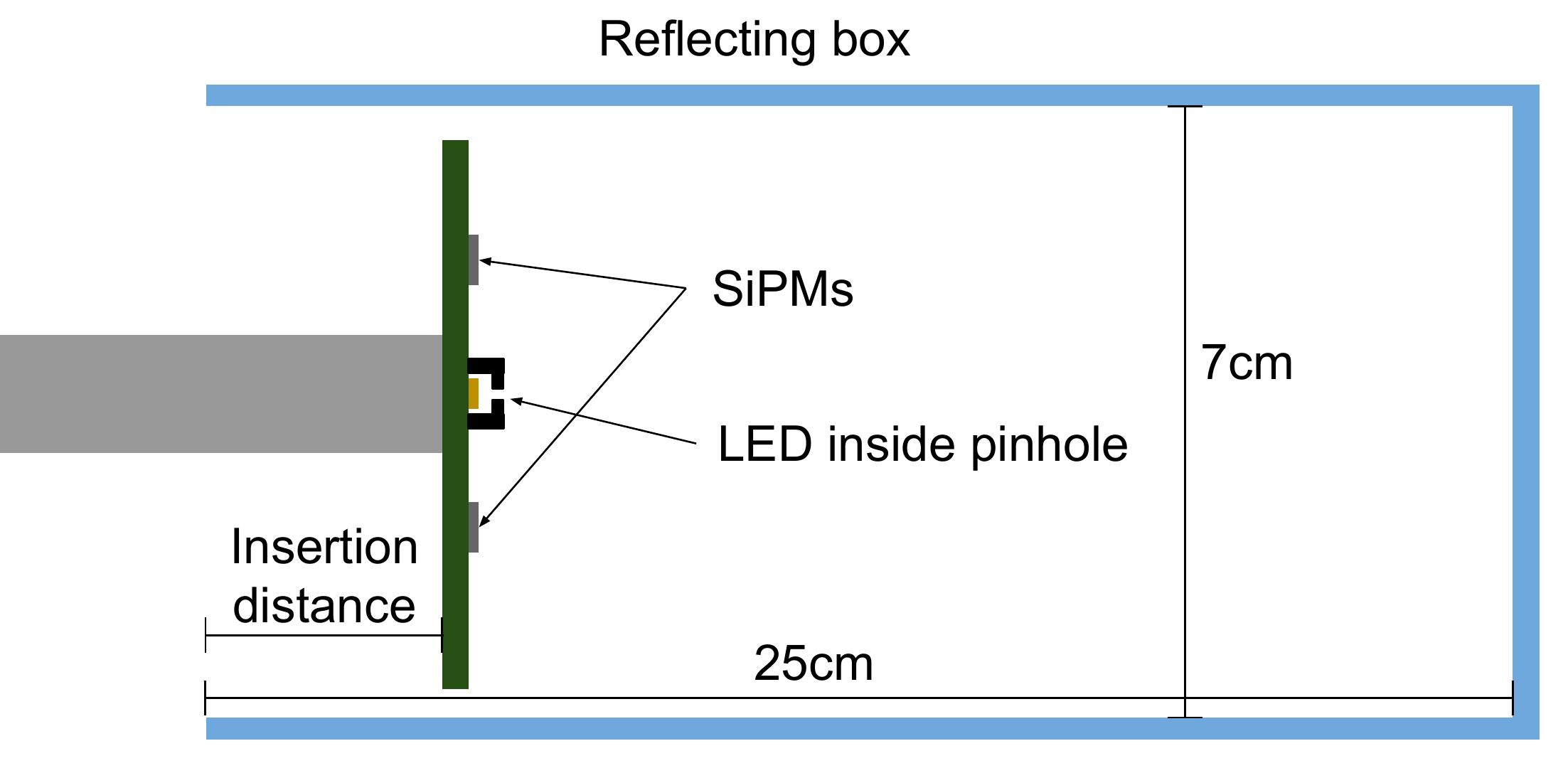}
\caption{Side-view schematic of the box setup. The inside dimensions of the box are $7~\mathrm{cm}~\times~7~\mathrm{cm}~\times~25~\mathrm{cm}$. The board holding the LED and SiPMs system is inserted through the open end of the box. The amount of light collected is measured as a function of the distance of the board insertion to extract the reflectance of the PTFE box.}
\label{fig:box_setup}
\end{figure}

The box method uses several cuboids with inner dimensions of $7~\mathrm{cm}\times7~\mathrm{cm}\times25~\mathrm{cm}$ made of PTFE pieces of a given thickness. The boxes are open at one end to allow the insertion of a printed circuit board with a light-emitting diode (LED) and 4 silicon photomultipliers (SiPMs). The board is mounted on a sliding post such that the insertion distance can be changed. Measurements of the light collected as a function of the distance of insertion are used to extract the reflectance. A schematic view of the setup is shown in Figure~\ref{fig:box_setup}. 

\begin{figure}[tb]
\centering
\includegraphics[width=.49\textwidth]{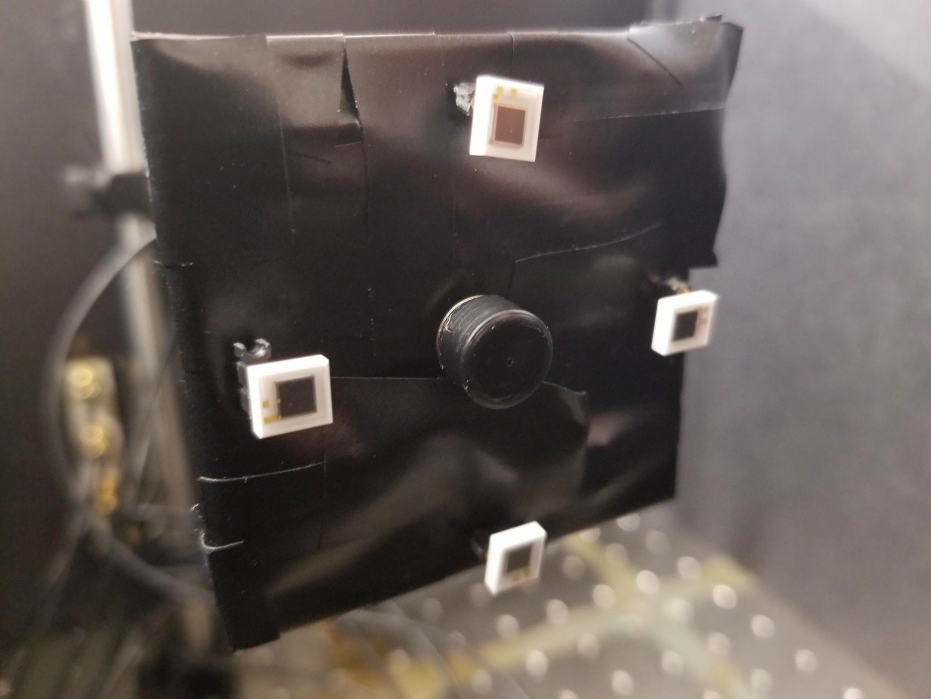}
\includegraphics[width=.49\textwidth]{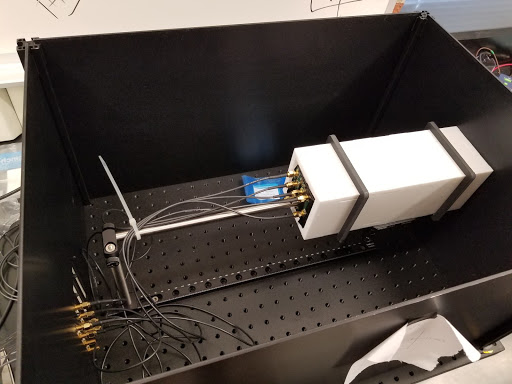}
\caption{Left: Picture of the LED-SiPMs system mounted on a printed circuit board. The LED, located in the center, is covered with a pinhole of 2~mm diameter. The black tape is intended to reduce reflection of light from the board. Right: Picture of the box setup with the readout board attached to a movable rod to allow variation of the insertion distance inside the box. The whole setup is inside a black box to ensure that external light does not affect the measurement.}
\label{fig:full_setup}
\end{figure}

Two different \textsc{ThorLabs} LEDs are used: an LED260W with light emission peaked at 260~nm, and an LED450LW with light emission peaked at 450~nm. These were chosen to be as close as possible to the light wavelengths of interest: the 175~nm VUV scintillation light of xenon, and the 420~nm TPB-shifted light. In all measurements, the LED is outfitted with a small pinhole made of opaque black plastic to collimate the light, reducing the importance of near-field effects and making the simulation of the setup easier. The SiPMs used for the main measurements are the $3\times3~\mathrm{mm
^2}$ UV-sensitive \textsc{Hamamatsu} S13370-3050CN. A different model, the $3\times3~\mathrm{mm
^2}$ \textsc{Hamamatsu} S13360-3050CS SiPMs, which are designed for visible light and offer no sensitivity to UV light, were also used to cross-check the measurements as both these SiPMs offer similar efficiency at blue wavelengths. The SiPMs are mounted at different locations on the board in order to average over any possible effects from non-uniformities in the PTFE, as shown in Figure~\ref{fig:full_setup}. Measurements are taken for insertion distances ranging from 0~cm to 21~cm at intervals of 3~cm. 

The pulses used to power the UV LED had a frequency of 50~Hz with a maximum voltage of 6.5~V and minimum voltage of 0~V. The ramp-up time was 5~ns and the pulse width was 200~ns. The pulses used to power the blue LED had a frequency of 50~Hz with a maximum voltage of 3.1~V and minimum voltage of 0~V. The ramp-up time was 5~ns and the pulse width was 200~ns. We based these characteristics on the stated tolerances of the LEDS in their product sheets\footnote{Blue LED: https://www.thorlabs.com/thorproduct.cfm?partnumber=LED450LW \newline UV LED: https://www.thorlabs.com/thorproduct.cfm?partnumber=LED260W}, and found that the LEDs performed stably under these parameters.

\begin{figure}[tb]
\centering
\includegraphics[width=.98\textwidth]{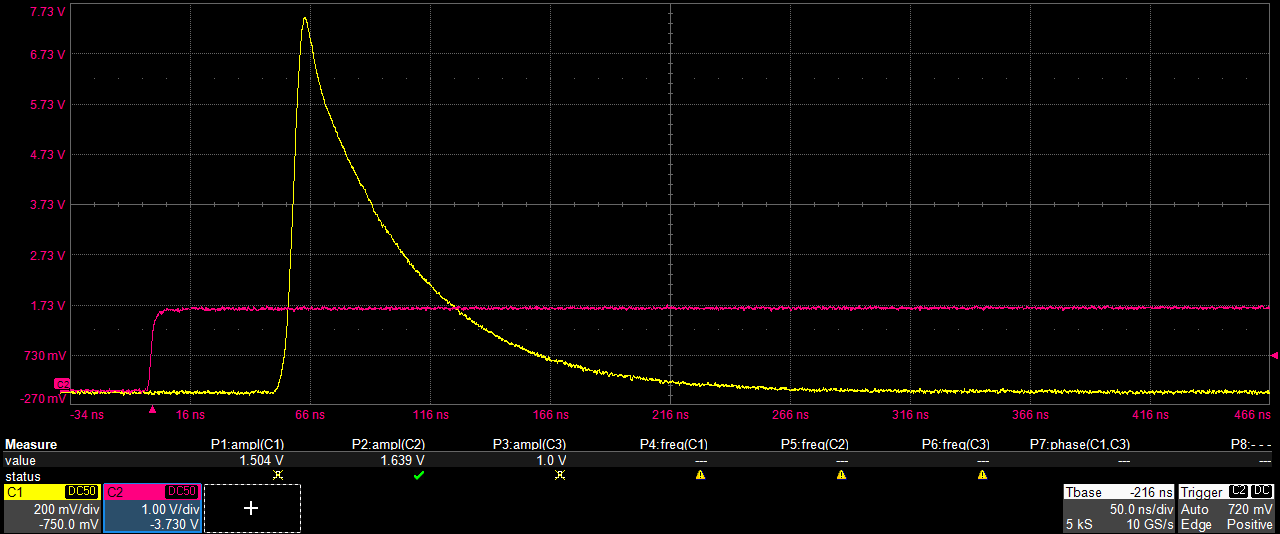}
\caption{Example SiPM pulse for box measurement. Yellow curve is SiPM response, pink curve is LED trigger.}
\label{fig:pulse_example}
\end{figure}

The signals from the SiPMs are directly fed into an oscilloscope. When measuring the signals from the SiPMs, about 1000 pulses are acquired at each position for each SiPM, and the average pulse area for all the pulses is calculated. A typical pulse is shown in Figure~\ref{fig:pulse_example}. A data point, as seen in Figure~\ref{fig:avg_sim_results}, is then obtained by summing the average areas for each SiPM. As a single bias voltage is applied in parallel across all four SiPMs, all detector signals were weighted by the inverse of their overvoltage. The SiPMs were provided with documentation indicating results of measurements of their individual breakdown voltages. The statistical error on each point is the standard deviation of the $\sim$1000 pulse areas divided by $\sqrt{1000}$. The statistical error was of order 0.01 V$\cdot$ns for all points. The systematic error is determined by repeating the measurement for each box four times, with the repeats spaced apart by an hour or so each. The systematic error was of order 1 V$\cdot$ns for all points.

\subsubsection{Extraction of the reflectance value}
In order to extract the reflectances from the box measurements, the data are compared with simulations of the setup for reflectances ranging from 80\% to 99.5\% in 0.5\% steps. The simulations are made using \textsc{Geant4} \cite{Allison:2016lfl}, where the box geometry includes the PTFE with a diffuse reflectance component only. Note that measurements with the spectrophotometer support the fact that any specular component to the reflectance of the PTFE is negligible, see Section~\ref{sec:foil}. The LED simulation, described in Appendix~\ref{appendix}, is dependant on the exact emission spectrum used. Therefore, in order to account for potential inaccuracies in the simulation of the LED emission spectrum, two different emission models, chosen to capture the most significant variation seen in the measured profile of the LED, were simulated, and the differences were used to assess a systematic uncertainty due to the LED emission profile. The statistical uncertainties of the individual models were calculated as $\sqrt{n (1-\frac{n}{N})}$, where $n$ is the number of photons detected in the simulation and $N$ is the total number of photons injected. These were added in quadrature and divided by 2 to determine the statistical uncertainty on the averaged simulation curve. In the blue LED with UV SiPM case, the uncertainty on the individual simulation curves was about 0.31 V$\cdot$ns, with the averaging contributing an additional $10^{-5}$ V$\cdot$ns statistical and 0.46 V$\cdot$ns systematic uncertainty. In the UV LED with UV SiPM case, the uncertainty on the individual simulation curves was about 0.72 V$\cdot$ns, with the averaging contributing an additional $10^{-4}$ V$\cdot$ns statistical and 0.93 V$\cdot$ns systematic uncertainty.  The outputs of the simulation are in the form of the number of photons emitted from the LED that reach the SiPMs, based on the different PTFE reflectances assumed. The simulation for the different reflectances was fitted to the data and the best fit was used as the reflectance value for that data set. More details on the fitting procedure are given below.

In order to allow comparison of the data taken with the SiPMs (in V$\cdot$ns) to the simulation (in photon counts), it is assumed that there is a constant conversion factor between the number of photons collected and the signal read out by the SiPMs. Therefore, the scale factor parameter ($\alpha$) is left free to vary in the fit. This $\alpha$ parameter is then the number of V$\cdot$ns expected per simulation photon, and soaks up effects such as the intensity of the LED and the efficiency of the SiPMs. The $\alpha$ scale factor should not change significantly between measurements of different reflectances, and the fit is performed simultaneously on all the data sets for the different reflectances using the same $\alpha$ parameter. Fig.~\ref{fig:avg_sim_results} illustrates how the fit of the different simulated reflectances is performed on the data set.

Although we expect the $\alpha$ value to be relatively stable, there is some possibility of variation coming in particular from the gain of the SiPMs changing in response to temperature. A cross check was run with a non-simultaneous fit allowing a different alpha value for each curve. Our results were consistent within error bars with those in Fig.~\ref{fig:avg_sim_results}.

\subsubsection{Results of the box method}
\begin{figure}[tb]
\centering
\includegraphics[width=.49\textwidth]{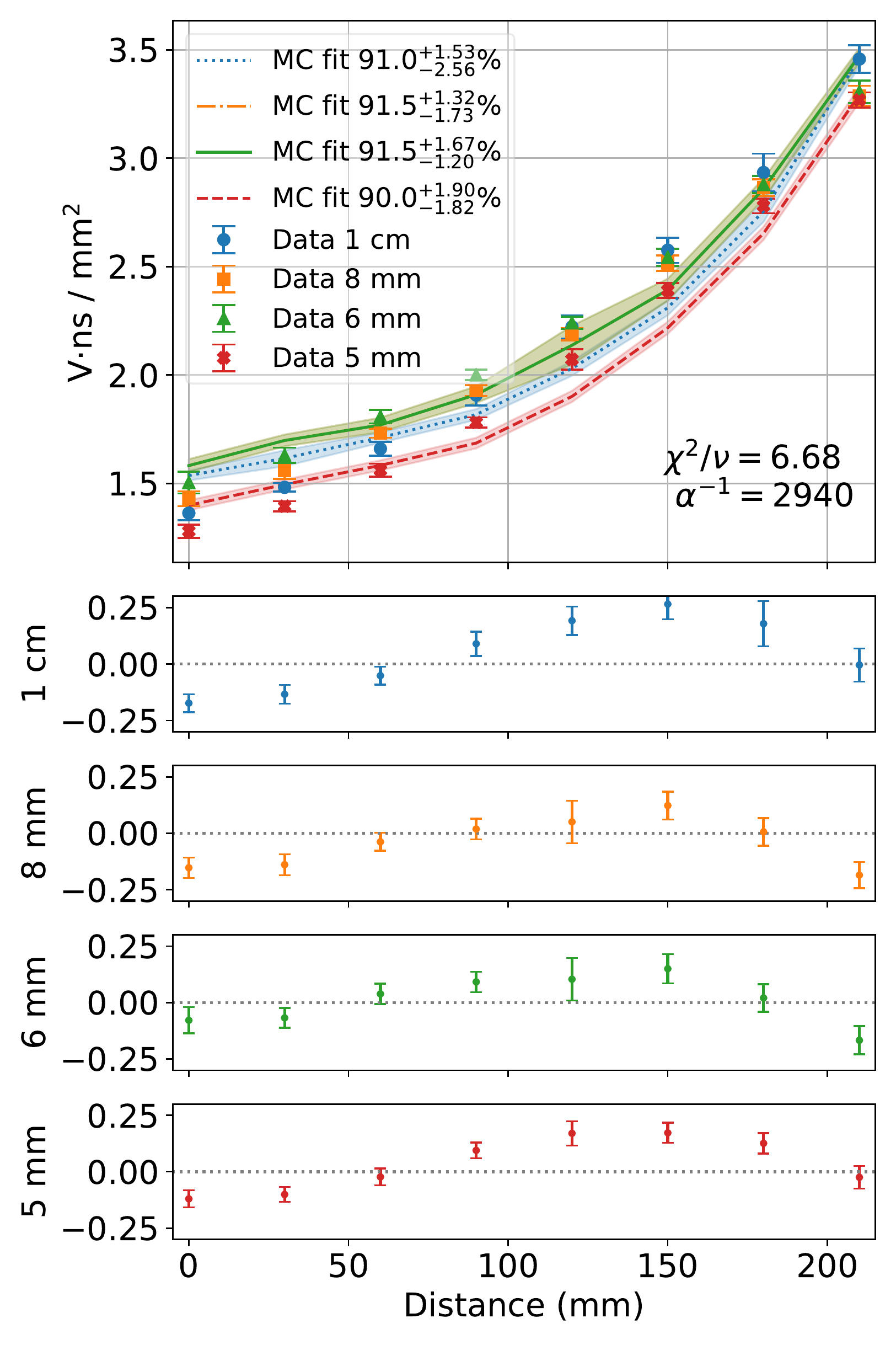}
\includegraphics[width=.49\textwidth]{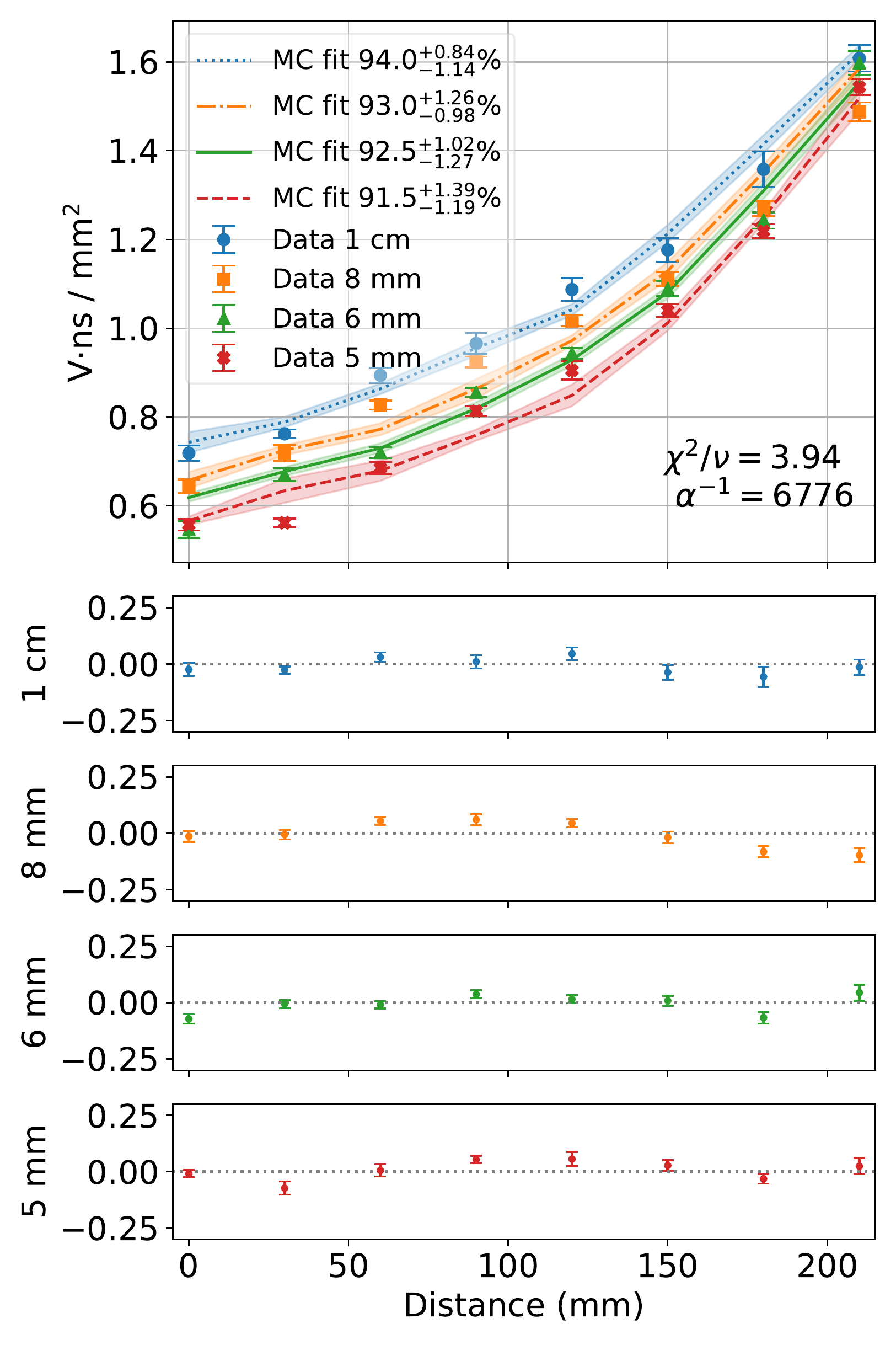}
\caption{Results for box measurements and fits. Upper row: The y-axis for the 260~nm UV LED with UV SiPM (left) and 450~nm blue LED with UV SiPM (right) gives the amount of light measured per unit photosensitive area of the SiPMs with pulse area measured in V$\cdot$ns. Error on the data points includes statistical uncertainties of the SiPM pulses and systematic uncertainties from repeating the measurement four times. Error on the simulations includes statistical uncertainties and systematic uncertainty evaluated by varying the LED emission profile using two distinct measurements. Lower row: These plots show the residuals from the fits to data in the upper row. The y-axis gives the difference between a data point and its corresponding simulated point (simulation minus data), and the x-axis gives the distances as in the upper row.}
\label{fig:avg_sim_results}
\end{figure}

Measurements of reflectances using the boxes were done for thicknesses of 5~mm, 6~mm, 8~mm, 10~mm. Figure~\ref{fig:avg_sim_results} shows the results obtained at 260~nm and at 450~nm. The simulation fit to the data provides the reflectance for each thickness.

For both plots in Fig.~\ref{fig:avg_sim_results}, it can be seen that the simulations reproduce the data relatively well. The agreement is worse for the UV LED results at small insertion, where the simulation produces a slower decrease in the number of photons observed. This could be explained by the effect of the simulated UV LED emission profile (see Appendix \ref{appendix}), which could affect results where the photons emitted at larger angles would have a greater chance to be absorbed when undergoing a higher number of reflections. This effect would be less important when the number of reflections is lower and when the insertion distance is larger.

For both wavelengths, the extracted reflectances of the PTFE samples are the same within error bars. This conclusion mostly comes from the large error bars from the fit of the simulations to the data, which could point towards inaccuracies in the simulations. The data points at 260~nm are close to each other, which is supported by the SPM measurements presented in Section~\ref{sec:spm_results_section} where the reflectance in the UV was the same within errors for all thicknesses. At 450~nm, the data points seem to indicate some separation between the different thicknesses. While the fit results do not allow quantitative conclusions, the data suggest a ~2\% variation between 5~mm and 10~mm.

\subsection{Comparison of the two methods}
\label{sec:comparison}

\begin{table}[tb]
\centering
\caption{Summary of PTFE reflectances at 260~nm and 450~nm for the box and photometer measurements. The photometer relative values are taken from Table \ref{tab:spm_main_result}, then normalized to the 10~mm~measurements from the box results for easy comparison. Error bars on the 10~mm~value from the photometer represent rescaled standard deviation between measurements.}
\label{tab:result_summary}
\begin{tabular}{ccccc}
\toprule
\multicolumn{1}{c}{} & \multicolumn{2}{c}{260 nm} & \multicolumn{2}{c}{450 nm}\\ 
\cmidrule(lr){2-3} \cmidrule(lr){4-5}
Thickness & Box & SPM & Box & SPM \\ 
\midrule
10 mm & 91.0\%$^{+1.6}_{-2.5}$ & 91.0 $\pm$ 0.4\% & 94.5\%$^{+0.9}_{-1.2}$ & 94.5 $\pm$ 0.2\% \\
8 mm & 91.5\%$^{+1.4}_{-1.6}$ & 91.6 $\pm$ 0.4\% & 94.0\%$^{+0.8}_{-1.1}$ & 94.4 $\pm$ 0.2\% \\
6 mm & 92.0\%$^{+1.3}_{-1.7}$ & 92.5 $\pm$ 1.3\% & 93.0\%$^{+1.1}_{-1.0}$ & 93.0 $\pm$ 0.2\% \\
5 mm & 90.0\%$^{+1.8}_{-1.7}$ & 91.2 $\pm$ 1.7\% & 92.5\%$^{+0.9}_{-1.5}$ & 91.1 $\pm$ 0.5\% \\
\bottomrule
\end{tabular}
\end{table}

Reflectances of PTFE were assessed with two different methods. The method using the SPM provided relative reflectances with smaller errors, and the box method fitted with simulations provided absolute reflectances with larger errors. In order to compare both methods, the photometer measurement values were scaled such that the results of the 10~mm thick PTFE pieces, used as the reference for the relative measurements of the SPM, matched with the 10~mm results of the box measurements. Table~\ref{tab:result_summary} shows the comparison of both methods, where the results are in good agreement with each other in both the UV and the blue.

\subsection{PTFE degradation} \label{sec:degrade}
\begin{figure}[!bt]
\centering
\includegraphics[width=0.49\linewidth]{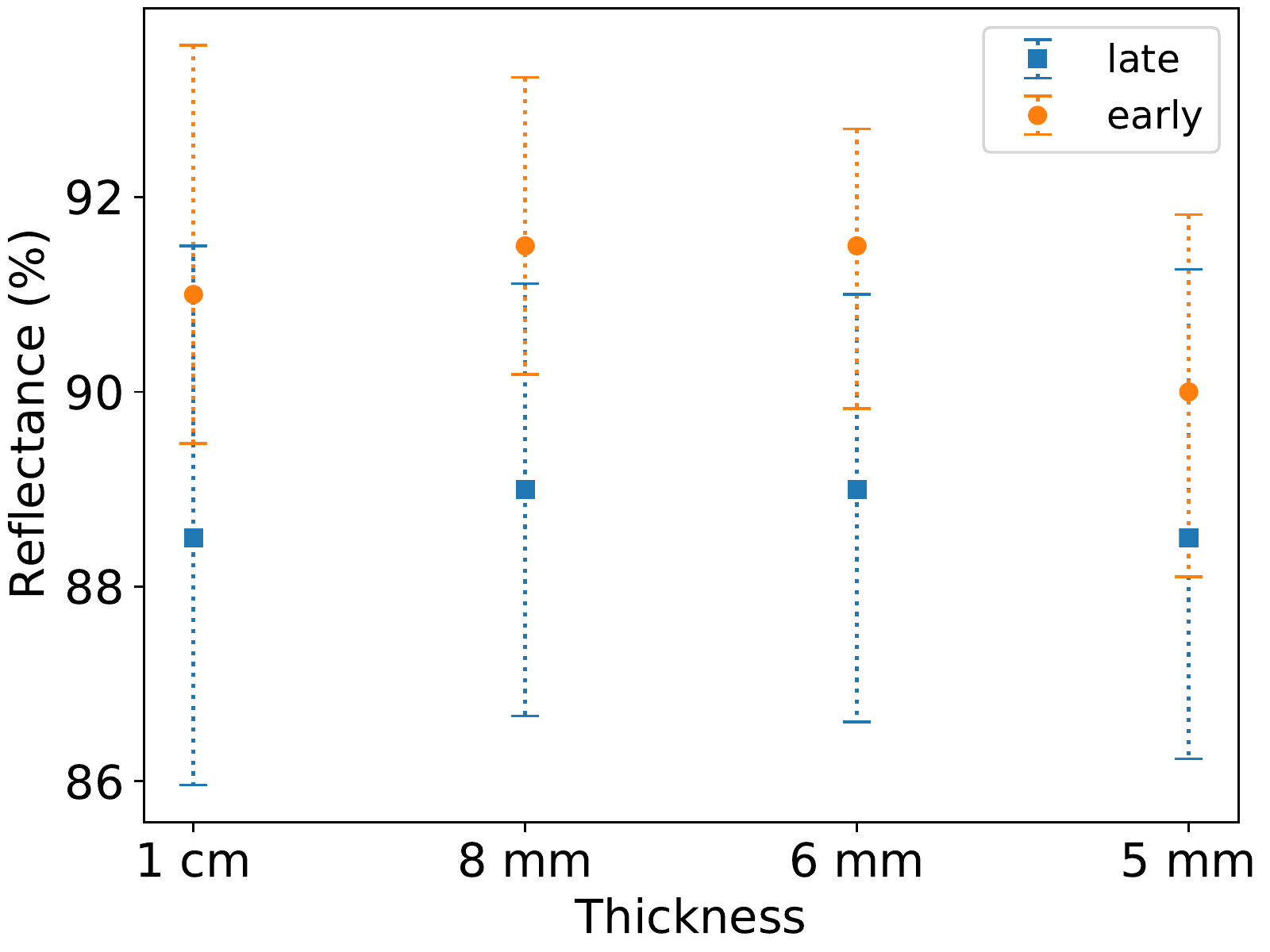}
\includegraphics[width=0.49\linewidth]{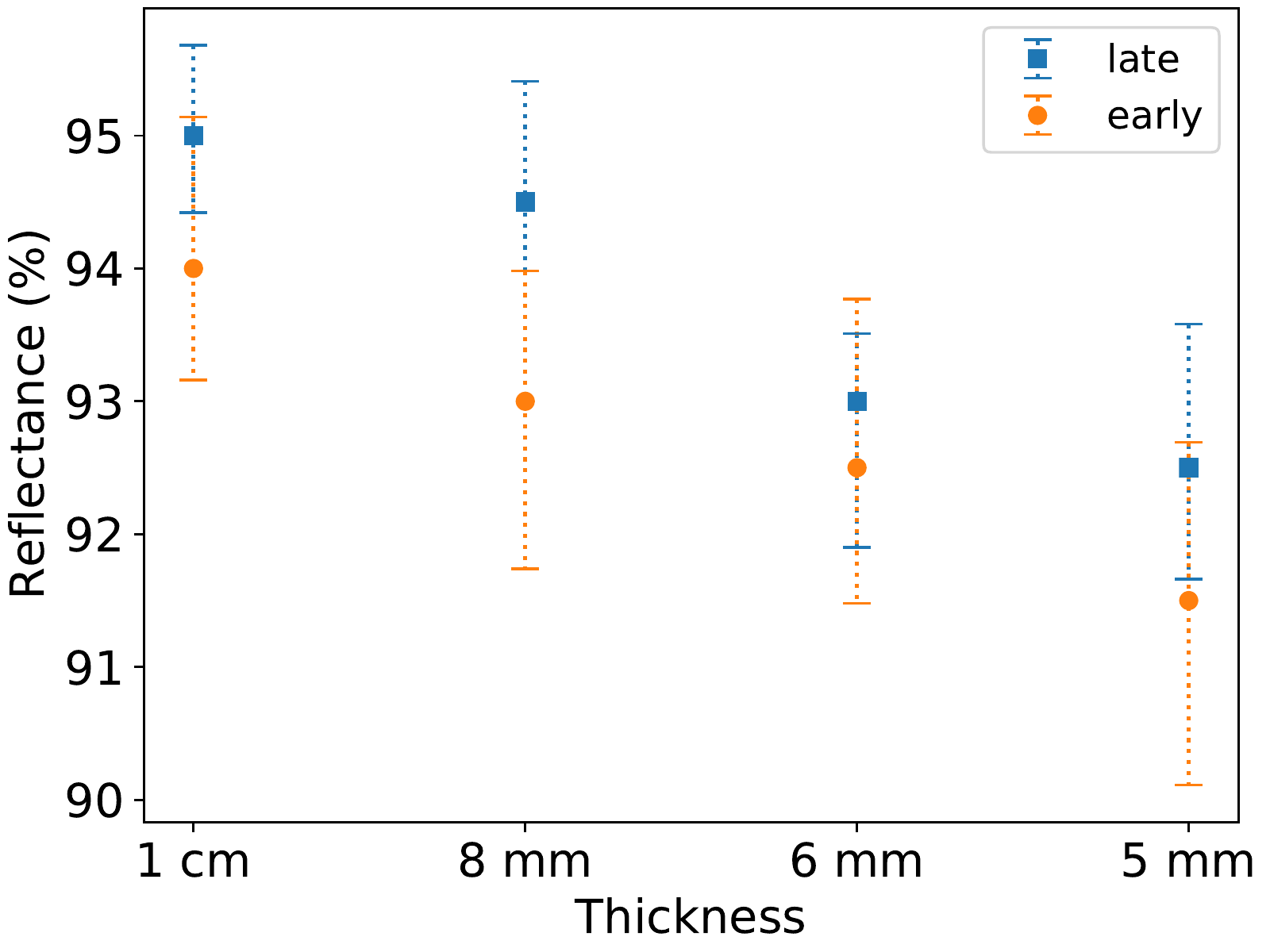}
\caption{Comparison of fit reflectances of the PTFE box measurements two months apart - "early" (circles) was taken in October 2019 and "late" (squares) was taken in December 2019. Left: Comparison for the UV LED/UV SiPM configuration. Right: Comparison for the blue LED/UV SiPM configuration. The early measurements were used for the results presented in the previous sections. Fit reflectances are all either within or nearly within uncertainties of each other. }
\label{fig:uvuv_degradation}
\end{figure}

The box reflectance measurements were repeated at a later date to investigate the possibility of reflectance variation over time. Figure~\ref{fig:uvuv_degradation} compares the results of these measurements to the original ones. While reflectances with the UV LED appeared to decrease slightly, reflectances with the blue LED were slightly increased. In both cases, however, the original and new fit reflectances are within uncertainty of one another. 

There was also a 3\% difference observed between the reflectances in the UV obtained using the SPM of the disks and of the box pieces. This may be attributed to the fact that while the boxes were kept covered when they were not being used, the PTFE disks were left exposed to light and likely degraded through UV exposure from ambient light. The timescale for this degradation was relatively short, around three weeks. Further investigation is required to confirm and quantify the degradation as the cause of this disagreement.

\section{Increasing PTFE reflectance with reflective foils} \label{sec:foil}
\begin{figure}[!tb]
\centering
\includegraphics[width=.45\textwidth]{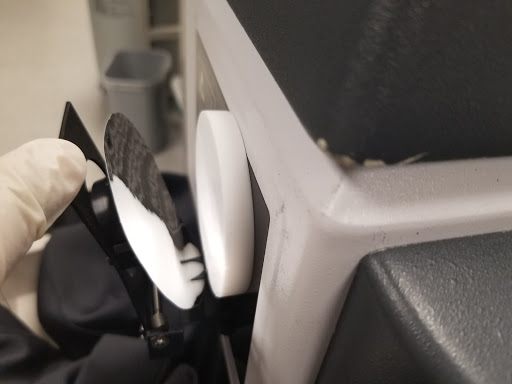}
\quad
\includegraphics[width=.45\textwidth]{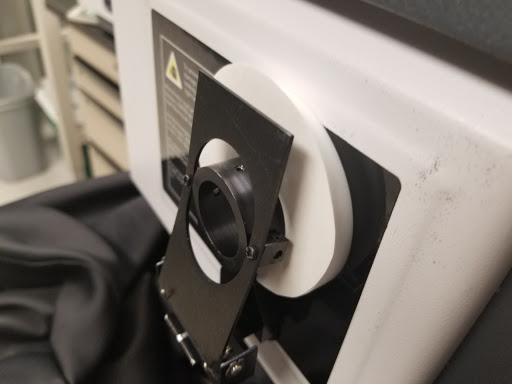}
\caption{Left: 3M foil being installed at the back of a disk of PTFE at the mouth of the integrating sphere. Right: 3M foil held on the back of a PTFE disk by the SPM lever arm.}
\label{fig:3M_setup}
\end{figure}

Although PTFE maintains good reflectance at relatively low thicknesses, it cannot be made arbitrarily thin and continue to function well as a reflector. It was investigated whether the mass of PTFE could be further reduced by introducing a backing specular reflector, without sacrificing reflectance. This possibility was addressed by looking at 3-inch diameter disks of PTFE backed with \textsc{3M} DF2000MA specular reflective foil. The foil is 38 microns thick. It was physically held in place by the same lever arm than holds the sample in place for the integrating sphere. This setup is shown in Figure
~\ref{fig:3M_setup}. There was interest in both the resulting total reflectance and the amount of specular reflectance. A specular component would represent different optical properties for PTFE.

In order to investigate lower thicknesses than were presented in the previous sections, two additional thicknesses of PTFE were sourced from \textsc{Nationwide Plastics}\footnote{www.nationwideplastics.net}. The main difference between the two PTFE providers is that the \textsc{Nationwide Plastics} sheets are skived from PTFE billets, while those from \textsc{ePlastics} are molded from PTFE resin. The two different types of PTFE are thus not directly comparable, but can still be compared to themselves before and after backing with foil.

The results of the 3M foil measurements are shown in Table~\ref{tab:3M_table}. A study to understand the potential introduction of a specular component to the diffuse PTFE reflector was also performed. The specular component (which should be 0 for a diffuse reflector) can be defined as the "diffuse excluded" component. In order to measure it, a measurement of the full reflectance is compared to a measurement where the specular reflection in the integrating sphere of the SPM is blocked out with black metal. This functionality is included with the integrating sphere. This gives the "diffuse" component of the reflectance; subtracting this component from the full measurement gives the specular component of the reflectance. In all cases, the specular component is zero within error bars. It was found that, at 450~nm, the relative reflectance of the foil (to the 1~cm disk) is 96.4 $\pm$ 3\%, whereas at 260~nm, it is 56.8 $\pm$ 1.4\%. The higher uncertainty relative to the other SPM measurements may be at least in part accounted for by the air gap between the foil and the PTFE disk, which may shift between setups.

\begin{table}
\centering
\caption{Comparison in spectrophotometer of reflectance at 260~nm and 450~nm between PTFE with and without 3M backing. All measurements relative to 10~mm (unbacked) PTFE disk. The total reflectance of the 3M foil alone at 450~nm (260~nm) is 101.6\% (40.7\%) relative to the 10~mm PTFE disk, and the specular component of the 3M foil alone is 76.9\% (16.2\%) relative to the 10~mm PTFE disk. Error bars come from standard deviation of repeated measurements of the same sample, done in triplicate, each time removing and replacing the sample. Samples from 5~mm to 10~mm are sourced from ePlastics, whereas 1.6~mm and 2.4~mm are Nationwide.}
\label{tab:3M_table}
\begin{tabular}{ccccc} 
\toprule
\multicolumn{1}{c}{} & \multicolumn{2}{c}{260 nm} & \multicolumn{2}{c}{450 nm}\\ 
\cmidrule(lr){2-3} \cmidrule(lr){4-5}
Thickness & PTFE & PTFE + foil & PTFE & PTFE + foil \\
\midrule
10 mm & 100\% & 99.9 $\pm$ 0.2\% & 100\% & 101.3 $\pm$ 0.2\% \\
8 mm & 96.7 $\pm$ 0.1\% & 96.8 $\pm$ 0.1\% & 98.8 $\pm$ 0.03\% & 100.3 $\pm$ 0.2\% \\
6 mm & 97.5 $\pm$ 0.2\% & 98.0 $\pm$ 0.1\% & 97.9 $\pm$ 0.1\% & 100.9 $\pm$ 0.2\% \\
5 mm & 95.9 $\pm$ 0.2\% & 96.8 $\pm$ 0.1\% & 96.0 $\pm$ 0.1\% & 100.6 $\pm$ 0.1\% \\
\hdashline
2.4 mm & 92.2 $\pm$ 0.5\% & 94.0 $\pm$ 1.1\% & 88.4 $\pm$ 0.2\% & 103.6 $\pm$ 0.6\% \\
1.6 mm & 87.4 $\pm$ 0.8\% & 89.2 $\pm$ 1.1\% & 84.6 $\pm$ 0.4\% & 100.3 $\pm$ 0.4\% \\
\bottomrule
\end{tabular}
\end{table}

In the blue, the results show a clear increase of reflectance --- even for 10~mm thick PTFE, backing with 3M gives a statistically significant improvement. Even more notably, it was observed that for thinner samples (2.4~mm and 1.6~mm), while the reflectance is normally low (< 90\%), after backing by foil, they are brought up to the same level of reflectance as the 10~mm PTFE. Furthermore, in no cases are there indications of a significant specular component, indicating that the optical properties of PTFE as a simple diffuse reflector remain.

In the UV, the foil is significantly less reflective, and thus provides a much smaller boost. Statistically significant gains are observed, which are larger for thinner PTFE, but never above a couple percent. In particular, none of the thinner PTFE is able to recover the reflectance of the 10~mm PTFE. The specular component remains negligible. The reflectance of 3M DF2000MA is between 10\% and 20\% at 260~nm, but it is also known to fluoresce when excited by UV light \cite{Geis:2017}. The relative importance of these two effects in our case would require further investigation.

\section{Conclusions and discussion}
Measurement of the variation in reflectance of PTFE with variations in thickness and wavelength (summarized in Table~\ref{tab:result_summary}) was presented as well as the impact of backing PTFE with specular reflective foil (summarized in Table~\ref{tab:3M_table}). It was found that the variation in reflectance at 450~nm is measurable, ranging from 92\% to 95\% for PTFE of thickness 5~mm to 10~mm at 450~nm. At 260~nm, the variation ranged from 91\% to 92\% but within errors for all thicknesses. A significant degradation of the PTFE reflectance in the UV is observed in PTFE disks, potentially due to exposure to ambient light, although more investigation is required to make a definitive and quantitative statement of the magnitude or timescale of cause and effect. There are slight changes in the reflectances of the boxes over time; likewise a more focused investigation is warranted. It also found that the reflectance of PTFE can vary up to 2.7\% when measuring different locations within a single sheet of material, making careful selection of PTFE important when constructing a detector. This also emphasizes the importance of having a method of measuring reflectance that can average over variations in reflectance over the surface of the PTFE, such as our box method. Note that an absolute difference of 2.7\% between two measurements gives a standard deviation of $2.7\% / \sqrt{2} = 1.9\%$, as seen in the measurement of reflectance at 260~nm and 5~mm in Table~\ref{tab:spm_main_result}.

Backing the PTFE with specular reflecting 3M foil gives considerable improvement at 450~nm, allowing even very thin PTFE to match the reflectance of 10~mm thick PTFE, without introducing a specular component into the reflectance spectrum. The improvement in the UV, however, is marginal.

As the challenges faced by the NEXT experiment are common to many rare event searches (large mass of PTFE for light collection, need for low background), we anticipate these results will be helpful widely to the community. In particular, while PTFE can vary from vendor to vendor, our initial results indicate that reducing thickness can be beneficial for room temperature experiments. While the results presented here are of high interest for experiments using noble element detectors, where PTFE is often used, the UV LED used had a longer wavelength than the scintillation light of Xenon (178~nm) and Argon (128~nm). A further study in detector medium will be performed in future work. In addition, the impact of the commonly used wavelength shifter TPB needs to be studied and will be investigated in a future work.

\acknowledgments
The NEXT Collaboration acknowledges support from the following agencies and institutions: the European Research Council (ERC) under the Advanced Grant 339787-NEXT; the European Union's Framework Programme for Research and Innovation Horizon 2020 (2014--2020) under the Grant Agreements No.\ 674896, 690575 and 740055; the Ministerio de Econom\'ia y Competitividad and the Ministerio de Ciencia, Innovaci\'on y Universidades of Spain under grants FIS2014-53371-C04, RTI2018-095979, the Severo Ochoa Program grants SEV-2014-0398 and CEX2018-000867-S, and the Mar\'ia de Maeztu Program MDM-2016-0692; the Generalitat Valenciana under grants PROMETEO/2016/120 and SEJI/2017/011; the Portuguese FCT under project PTDC/FIS-NUC/2525/2014 and under projects UID/04559/2020 to fund the activities of LIBPhys-UC; the U.S.\ Department of Energy under contracts No.\ DE-AC02-06CH11357 (Argonne National Laboratory), DE-AC02-07CH11359 (Fermi National Accelerator Laboratory), DE-FG02-13ER42020 (Texas A\&M) and DE-SC0019223 / DE-SC0019054 (University of Texas at Arlington); and the University of Texas at Arlington (USA). DGD acknowledges Ramon y Cajal program (Spain) under contract number RYC-2015-18820. JM-A acknowledges support from Fundaci\'on Bancaria ``la Caixa'' (ID 100010434), grant code LCF/BQ/PI19/11690012. Finally, we thank Brendon Bullard, Paolo Giromini and Neeraj Tata for helpful discussions and assistance with preliminary measurements.

\appendix
\section{Simulating the LED}
\label{appendix}

It was found that simulating the LED as a point source fit the data poorly. To account for an extended source, at least 2 intensity profiles at different positions were required, as any single profile could be produced by a point source. The LED was measured at two downstream locations using a beam profiler (\textsc{Thorlabs BP209-VIS}), one near the LED and one far. The blue LED was measured at 0.1 and 0.3 inches from the face of a ThorLabs beam profiler. The UV LED was measured at 0.1 and 0.2 inches from the face of the profiler. The profiles measured were 1 dimensional in x and y, with the overall profile taken as the product of the x and y profiles.

\begin{figure}[!htb]
\centering
\includegraphics[width=0.9\textwidth]{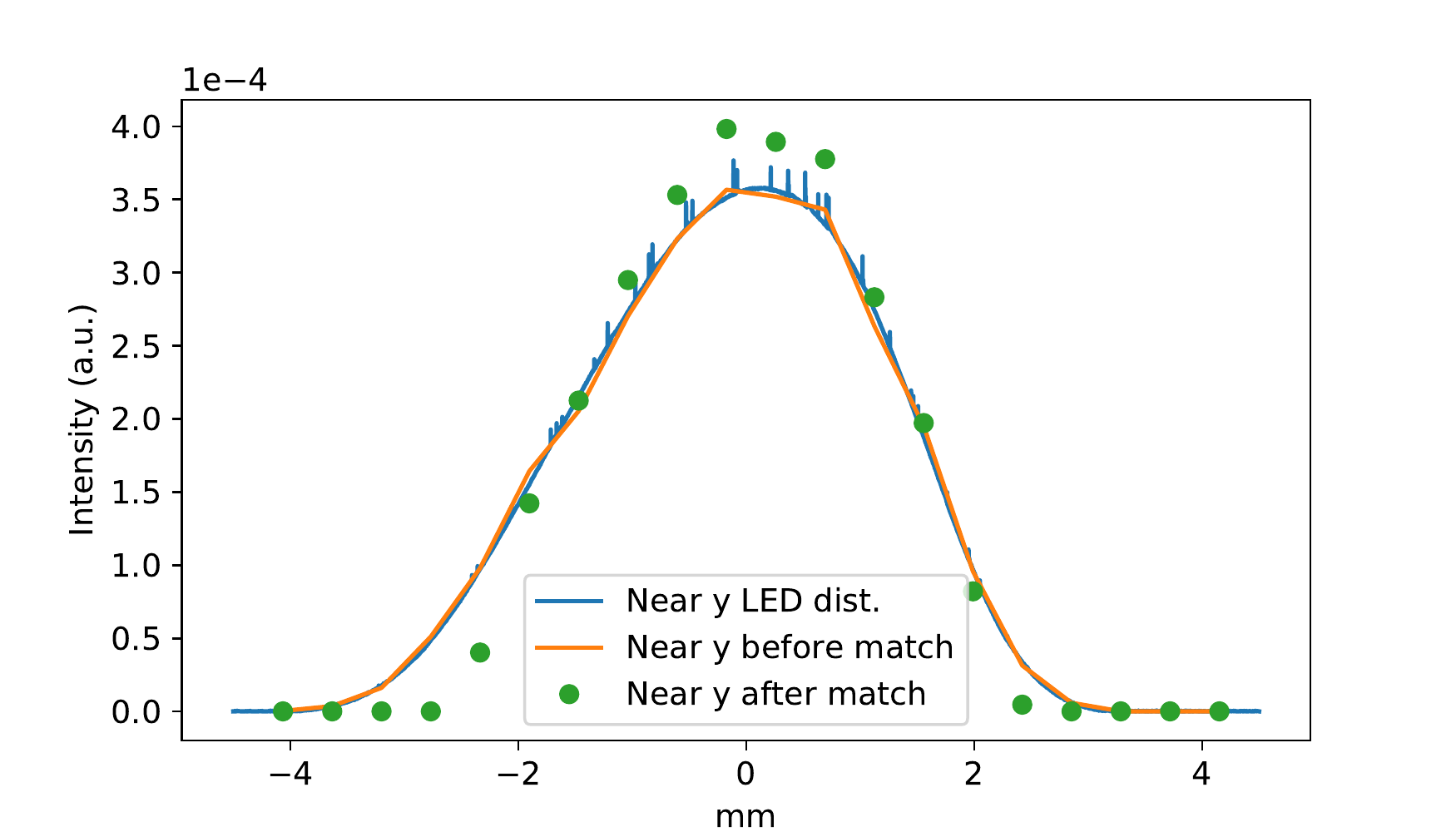}
\caption{Near y profile simulations with 10,000 hits. The generated distribution before matching is included to indicate that some of the failure to match the empirical distribution comes from fluctuations in the generation of the hits on each plane.}
\label{fig:near_y}
\end{figure}

\begin{figure}[tb]
\centering
\includegraphics[width=.475\textwidth]{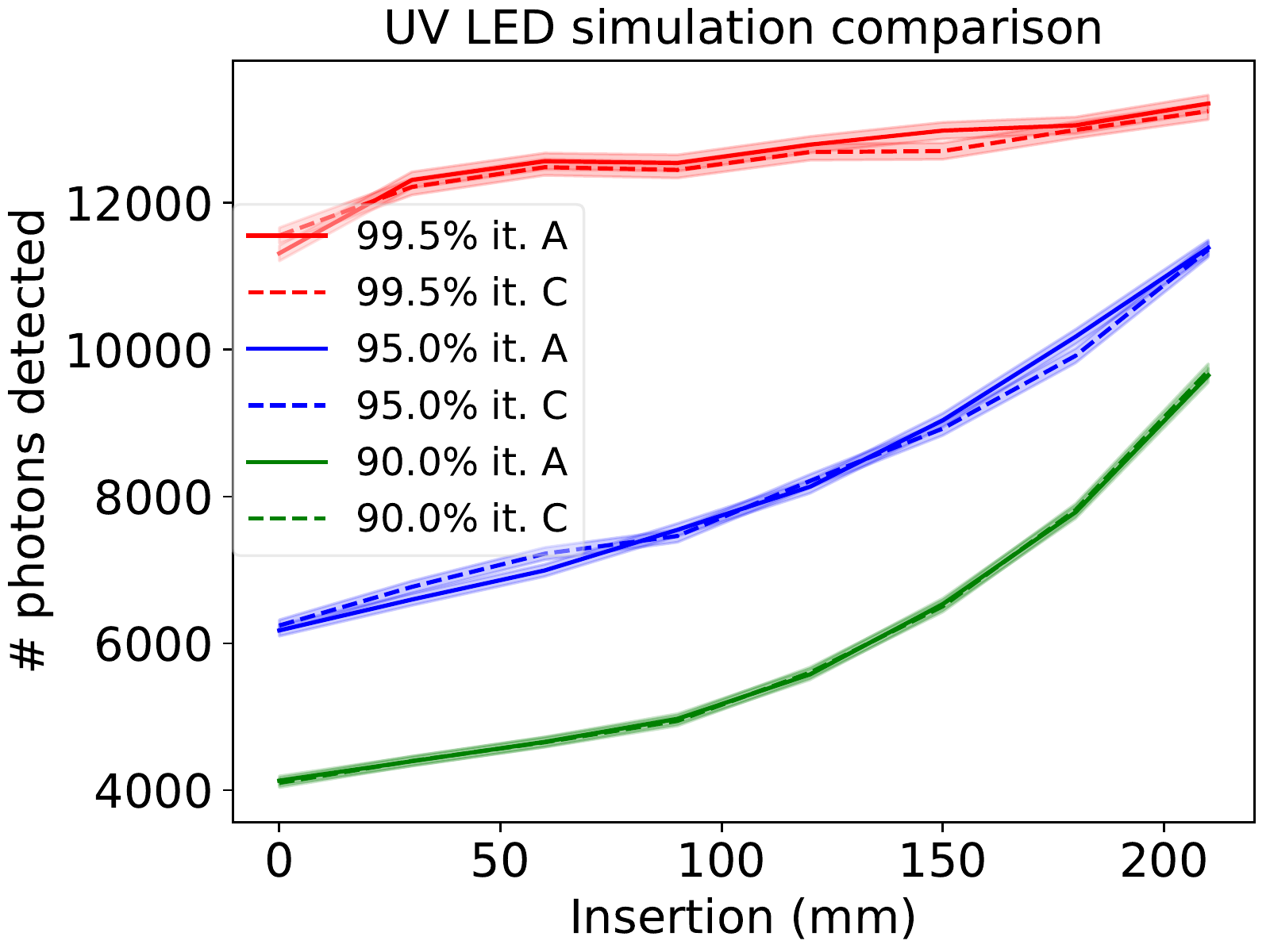} \quad
\includegraphics[width=.475\textwidth]{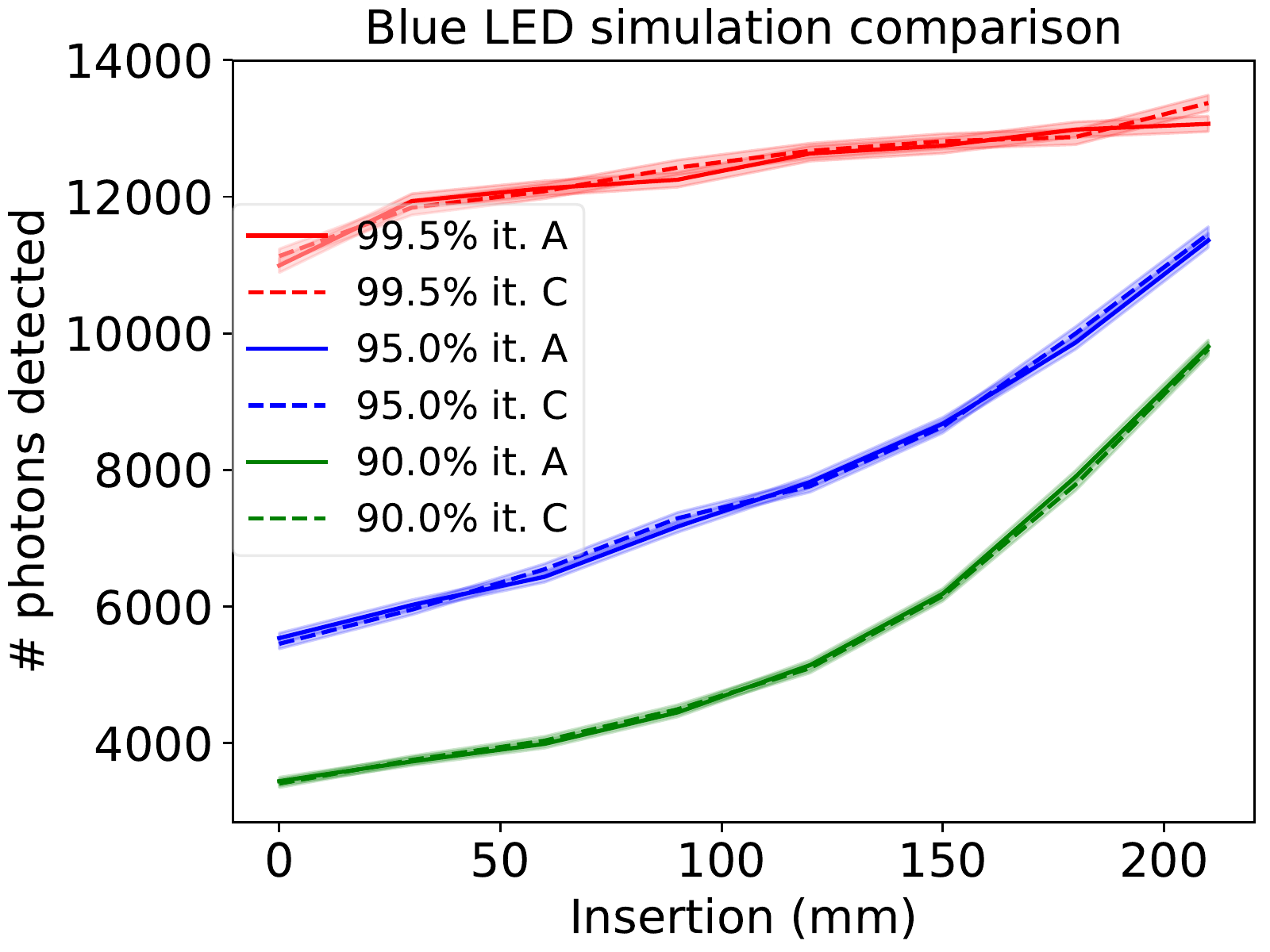}
\caption{Comparison of simulation curves obtained with iteration A and iteration C LED profile measurements for  the UV LED (left) and the blue LED (right). The changes in the profiles do not significantly impact the simulation results.}
\label{fig:simulation_iteration_comparisons}
\end{figure}

In order to generate simulation photons, 10,000 hits were generated on both the near and far plane, according to the measured distributions. Hits were then matched, where a "match" consists of a near hit and a far hit where the line between them lands correctly on the source. Some amount of distortion arises from the fact that generally not all hits will be able to find a match. However, this number of unmatched hits is reasonably small (10\% for 10,000 hits), and thus the output profiles are reasonably good matches to the physical LED.

To check the effectiveness of this procedure, the output profile at the near and far planes was compared to the measured profile by taking the RMS difference point to point between the respective histograms and scaling relative to the maximum height of the profile. Doing so gave around 5\% disagreement in the near x and y profiles, and 3\% difference in the far profiles. This held regardless of the LED being simulated.

Another cross check was to measure a third profile, then use photons generated from the first two profiles to try to recover the information on the third ("interpolated") profile. This profile sat between the other two, at 0.2 in in the blue case, and 0.15 in the UV. No information on the interpolated profile is input in the simulation. The disagreement between the interpolated profiles was, in all cases, around 10\%. A sample of the near profiles can be seen in Figures~\ref{fig:near_y}.

In order to understand the possible impact on reflectance fits, the profile measurements were repeated for four iterations (labeled A through D). These profiles were visually inspected for both LEDs to see how they differed, and then a pair was chosen (iteration A and iteration C) which maximally differed from one another. One set of simulations was run using each, and the results were compared (Fig.~\ref{fig:simulation_iteration_comparisons}). As the simulations did not appear to be significantly impacted by the change in profiles, the box fits were performed using an average of the iteration A and iteration C simulation curves, propagating those error bars into the uncertainties.


\providecommand{\href}[2]{#2}\begingroup\raggedright\endgroup

\bibliographystyle{JHEP}

\end{document}